\documentclass{an}
\usepackage{graphicx}
\usepackage{times}
\Volume{00}              
\Year{0000}              
\Month{00}               
\Pagespan{000}{000}      

\begin{document}
\lhead[\thepage]{A.N. S.F.S\'anchez: The Euro3D Visualization Tool}
\rhead[Astron. Nachr./AN~{\bf XXX} (200X) X]{\thepage}
\headnote{Astron. Nachr./AN {\bf 32X} (200X) X, XXX--XXX}

\title{E3D, the Euro3D Visualization Tool II: \\
mosaics, VIMOS data and large IFUs of the future.}

\author{S.F. S\'anchez\inst{1} \and T.Becker\inst{1} \and A.Kelz\inst{1}}
\institute{Astrophysikalisches Institut Potsdam, And der Sternwarte 16, 14482
  Potsdam, Germany }
\date{Received {date will be inserted by the editor};
accepted {date will be inserted by the editor}}

\abstract{ In this paper, we describe the capabilities of E3D, the Euro3D
  visualization tool, to handle and display data created by large Integral
  Field Units (IFUs) and by mosaics consisting of multiple pointings.  The
  reliability of the software has been tested with real data, originating from
  the PMAS instrument in mosaic mode and from the VIMOS instrument, which
  features the largest IFU currently available.  The capabilities and
  limitations of the current software are examined in view of future large
  IFUs, which will produce extremely large datasets.
\keywords{techniques: image processing; methods: data analysis }
}
\correspondence{ssanchez@aip.de}

\maketitle

\begin{figure*}
\resizebox{\hsize}{!}
{\includegraphics[width=1.0\textwidth]{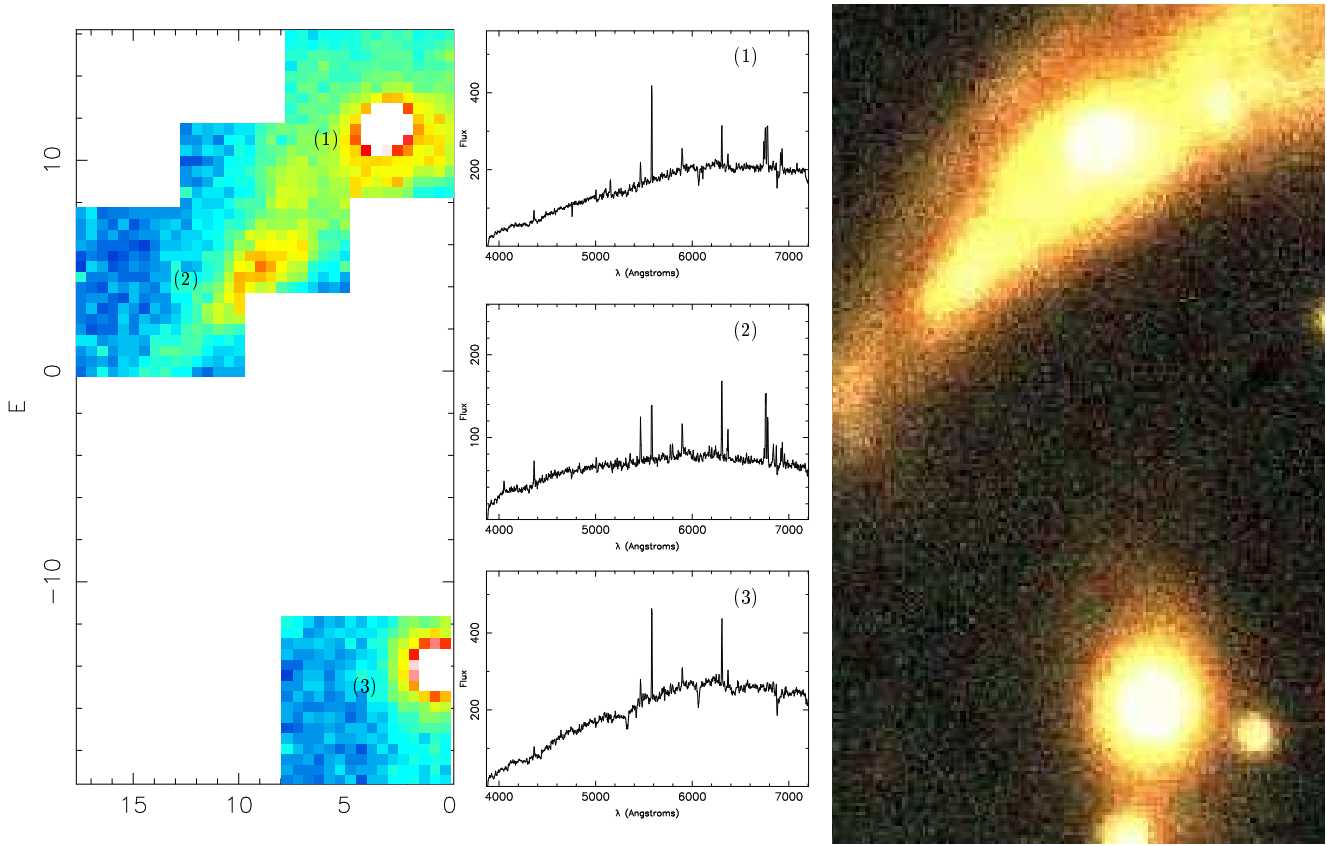}}
\caption{ Left: re-constructed H$\alpha$ maps (FWHM=10~\AA) of four mosaic
  pointings (each FoV = $8''\times8''$) of IRAS 16365+4202, observed with
  PMAS.  Center: Selected spectra of regions as indicated on the maps, on
  arbitrary units. Right: true-color image of the same region of the objects,
  created using BVR images acquired with the PMAS A\&G before the acquisition
  of the spectra.}
\label{mosaic}
\end{figure*}

\section{Introduction}

In S\'anchez (2003), hereafter Paper I, we presented E3D, the Euro3D
Visualization tool. We described its main characteristics and gave some
examples of its use with different IFU data. One of the initial requirements
for the Euro3D visualization tool was that it shall be able to handle large 3D
datasets made either from large number of exposures of the same ``small'' IFU
at different pointings ({\it mosaics}) or from a single or few
exposures obtained using a ``large'' IFU. To demonstrate the capabilities of
the current software to handle and display mosaics and large datasets, we give
examples of its use for the visualization of a mosaic of PMAS data
(\cite{san03}), section 2, and a single exposure obtained with the largest
existing IFU, namely VIMOS/IFU (\cite{san03}), section 3. We then discuss the
current limitations of the E3D software in view of future very large IFUs like
the MUSE/VLT project (\cite{bacon2002}).

\begin{figure*}
\resizebox{\hsize}{!}
{\includegraphics[width=1.0\textwidth,viewport=20 300 580 820,clip]{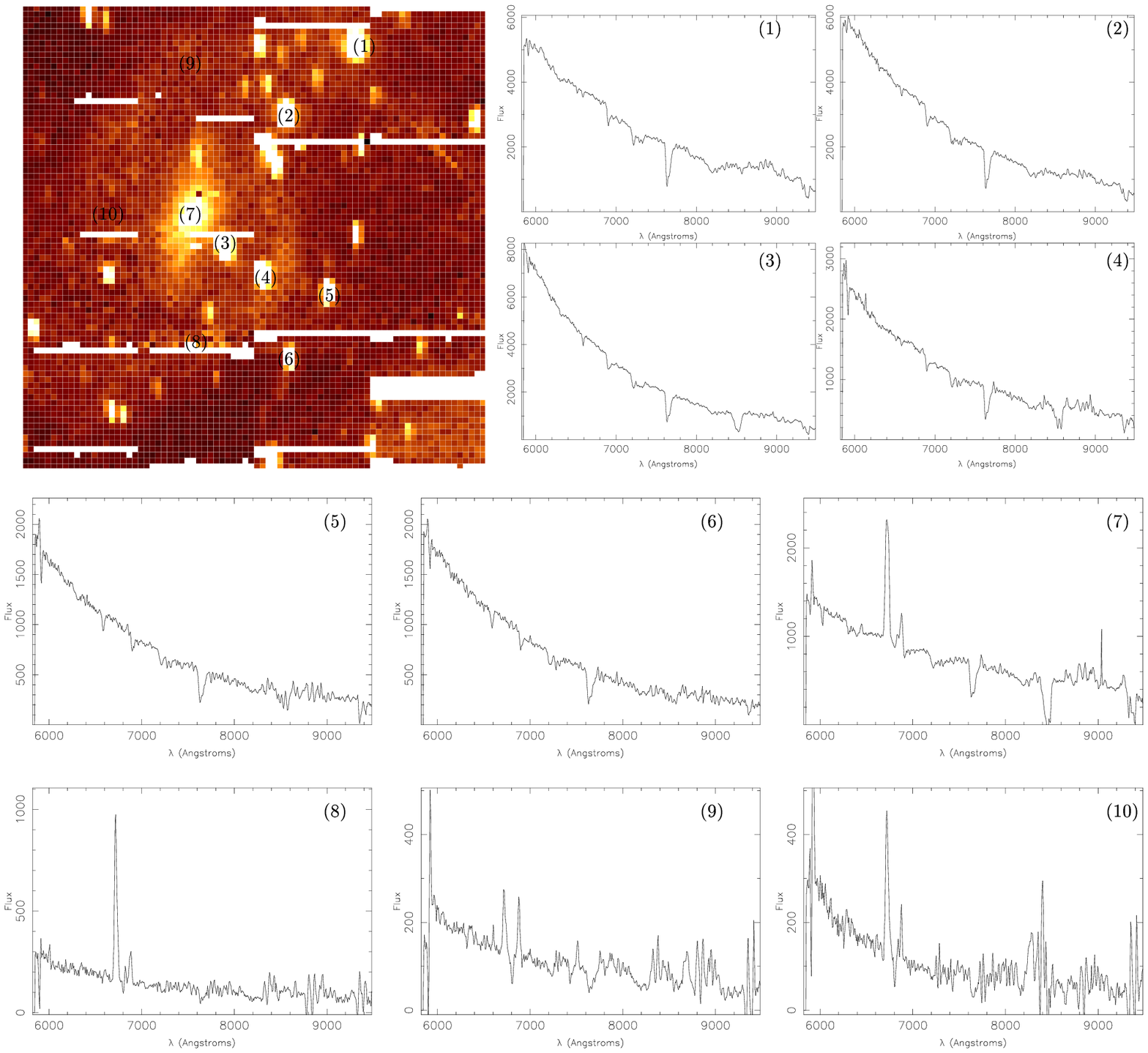}}
\caption{Upper-left pannel: re-constructed $H_{\alpha}$ image of IRAS 13031-5717, resulting from  VIMOS data that has been reduced with P3d (\cite{be01}) and cleaned with E3D
  (S\'anchez 2003) (for further explanation see text).  Other pannels: 
  spectra of selected areas as indicated in the field, i.e.  bright foreground
  stars (no. 1-6) and different regions of the target (no. 7-10) (Note: the
  white strips in the image corresponds to regions of low-performance
  spaxels, which have been removed during the cleaning process).}
\label{example}
\end{figure*}

\section{Mosaics}
\label{sec:mosaic}

Figure \ref{mosaic} shows a mosaic of four pointings, obtained with the PMAS
instrument (\cite{ke03}, \cite{ro00}) of the object IRAS 16365+4202.  Three of
the pointings overlapp (with 6 $\times$ 8 coincident spaxels each, where a
{\it spaxel} is defined as a spatial element: fiber, lenslet...) while one
pointing is completely separated spatially (centered on a nearby companion).
The size of each individual field of view is $8'' \times 8''$.  The mosaic of
the overlapping pointings has been created using an IDL based software code
written by one of us (\cite{be01}), while the single, non-contiguous, pointing
was included using the standalone program {\tt pmasMosaic}, that is part of
the E3D distribution.

Both the Euro3D format and E3D handle complex geometries of mosaic pointings,
without the need to add {\it fake} values or to fill any gaps, like in the
case of a 3D cube, or to interpolate the data to match a certain grid or
pattern. This method decreases the size of the stored data on the disk and
increases the speed of reading/writting. Importantly, it also preserves the
original data (as opposed to interpolations). The mosaic displayed in In Fig.
\ref{mosaic} comprises 934 spectra with 1009 spectral pixels each one. Each
pixel contains two times a float digits of 4 bytes, one for the data, and one
for the noise information. i.e., 8 bytes per spectral pixel and spaxel. The
final size of this file on the disk is $\sim$7.3Mb. It takes less than 3
seconds to E3D to load the complete file in the memory, using a Linux PC with
512Mb of RAM and a Pentium IV processor at 2GHz (hereafter {\tt C1}), and
about the double using a Linux PC with 128Mb of RAM and a Pentium III
processor at 700MHz (hereafter {\tt C2}). If the data were storaged on a
continuously sample datacube (i.e., in which all the gaps between the
pointings are filled with ``fake'' data), a cube containing more than 2520
spectra would be needed, yielding a total size of 20Mb on the disk.

\section{Large IFUs: test with VIMOS}
\label{sec:largeIFU}

We have tested E3D with low-resolution VIMOS (\cite {lef03}) data. These data
were obtained for an on-going project dedicated the study of the connection
between merging and AGN activity in galaxies (\cite{san03}). They reduced
using a modified version of the PMAS data reduction software P3d
(\cite{be01}). VIMOS, mounted on the VLT, is the largest existing IFU. It
comprises 6400 spaxels, with 536 spectral pixels each (in the low resolution
mode). A single VIMOS exposure occupies $\sim$26.5Mbs on the disk, saved in
the Euro3D format. E3D reads this file in $\sim$7 seconds, using {\tt C1}, and
about the $\sim$17 seconds using {\tt C2}.

VIMOS has a number ($\sim$10\%) of dark or low-performance spaxels, that are
not homogeneously distributed within the field-of-view. We have used E3D to
eliminate these spaxels. As a consequence of the Euro3D format, dead and/or
low-efficiency spaxels can be easily removed from the data, without affecting
the data visualization. In Figure \ref{example} we present an example of this
cleaning method. This figure shows an H$\alpha$ map of the object IRAS
13031-5717, observed with VIMOS in the low-resolution red mode. All the bright
point-like objects on the field are foreground stars. We also plotted the
spectra of foreground stars (spectra 1 to 6) and different regions of the
target (spectra 7 to 10). The removed spaxels appear as white strips in the
image.

E3D includes 5 different interpolation algorithms: Spline, Linear Delaunay,
Natural Neighbors, Linear Nearest Neighbors and Inverse Distance Nearest
Neighbors. The fastest method is the Linear Delaunay, although the most
accurate is the Natural Neighbor. The speed of an interpolation depends on the
original number of spaxels, and the final number of pixels of the regular
grid. It takes less than $\sim$0.1 seconds to E3D to interpolate a VIMOS map
(6400 spaxels) in a regular grid of 0.2$\arcsec$$\times$0.2$\arcsec$ pixels
(90000 pixels) using the Delaunay interpolation, and $\sim$1 second using the
Natural Neighbor interpolation (for {\tt C1} configuration).

\begin{figure*}
\resizebox{\hsize}{!}
{\includegraphics[width=1.0\textwidth]{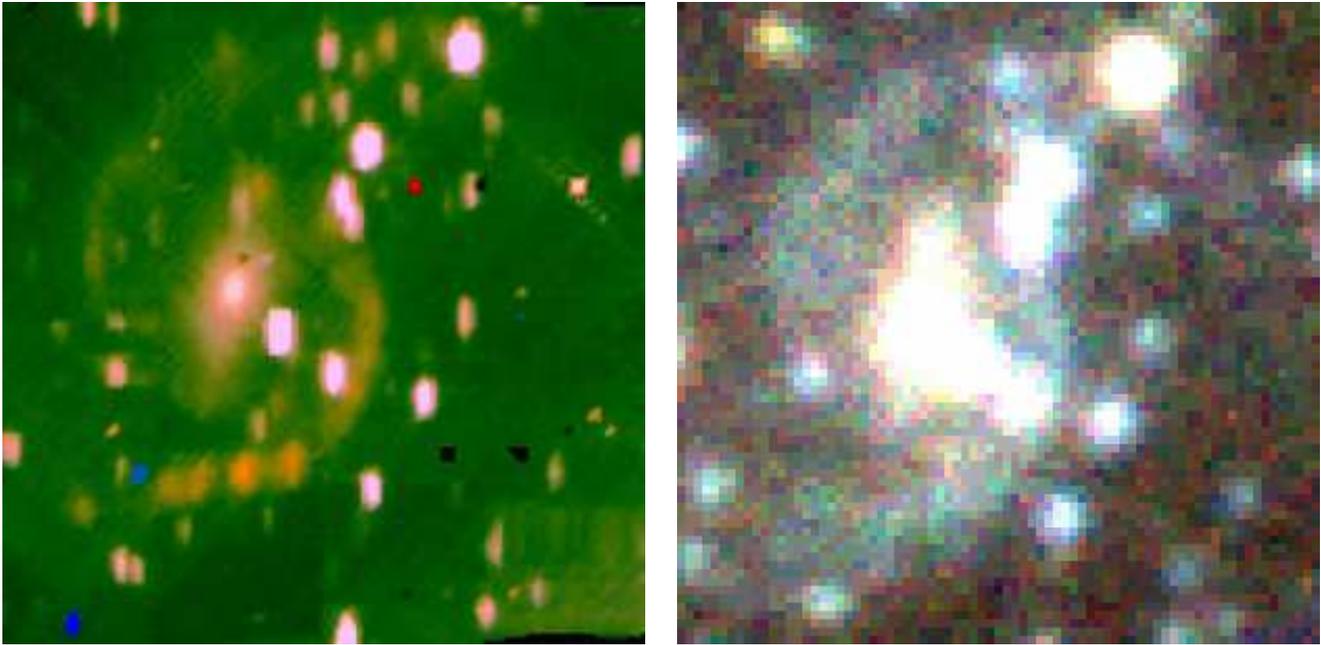}}
\caption{Left pannel: True-color image of IRAS 13031-5717.
   The image was created by combining three polychromatic maps (maps created
   by integration of a given spectral range) of cleaned and
   interpolated VIMOS data, centered at H$\alpha$ and the continuum on both
   sides. Right pannel: True-color image of the same object created using the
   $B$,$R$ and $I$-band images from the SuperCOSMOS catalogue.}
\label{color_map}
\end{figure*}

We have interpolated the VIMOS cleaned data, using the Linear Delaunay
algorithm, and created three images, that have been combined using the EXPORT
routine under IRAF to generate a true-color image. The three wavelength ranges
used for the true-color image of Figure \ref{color_map} are the H$\alpha$
emission line and the continuum bluewards and the continuum redwards of it. For
comparison, we have included a true-color image of the same field obtained
using the EXPORT routine over $B$, $R$ and $I$-band images from the SuperCOSMOS
catalogue (\cite{ha01}).  Comparing Figures \ref{example} and both images on
Figure \ref{color_map} we can appreciate that the interpolation routine works
well in general. There are some artifacts created by the interpolation on the
areas with large strips of missed data. The artifacts are clearly identified
as purely red, blue or black spots, and can be easily removed by hand. We have
prefered not to clean them to show the limits of the interpolation over large
datasets with strips of non existing data.

\section{Future work}
\label{sec:future}

A problem with the current software version is that it does not deallocate a
fraction of the memory in the process of reading large data formats.  For
example, using {\tt C1}, the datasize limit is reached at $\sim$64000 spaxels
(assuming spectra of $\sim$1024 spectral pixels, and float digits) or any
comparable spectral/spatial combination.  Under these conditions it takes
about 3 minutes to read the data.
This shortcoming will be a problem for the huge datasers of possible future
instruments like MUSE (\cite{ba02}), which will have more than 90000 spaxels,
each containing 4k spectral points. It will therefore have to be fixed. It is
however worth pointing out that current instruments like VIMOS (6400 spaxels),
or reasonable mosaics (e.g. up to 250 pointings with a PMAS-like IFU) are not
affected by this problem.

\acknowledgements

This project has be founded by the Euro3D Training Network on Integral Field
Spectroscopy, funded by the European Commission under contract No.
HPRN-CT-2002-00305.

I'd like to acknowledge Sebasti\`en Foucaud for his help on the development of
an interface to comunicate between E3D and external packages, especially
with PYTHON. I like to acknowledge Matthew Horrobin for his tests with SPIFFI
and his useful comments. I like to acknowledge A.P\'econtal-Rousset, P.
Ferruit and all the Lyon group for their advice, help, and their marvelous
work on the library. Thanks to all the network for your comments and help.

I'd like to acknowledge the referee, P.Ferruit, for his comments that
have help to improve substantially the quality of this article.



\begin{thebibliography}{}
\bibitem[Bacon et al.\ 2002] {bacon2002} Bacon, R., et al., \ 2002, in
   {\it Scientific Drivers for ESO Future VLT/VLTI Instrumentation},
   eds. J. Bergeron, G. Monnet, Springer, Berlin, p.~108
\bibitem[Becker 2001]{be01} Becker, T., 2002, PhD Thesis, University of Potsdam, Germany
\bibitem[Becker et al. 2003]{be03} Becker, T., Roth, M.M., Kelz, A., 2003, Euro3D Science Workshop, 21-23 May 2003, IoA, Cambridge, AN, , these proceedings.
\bibitem[Hambly et al. 2001]{ha01} Hambly, N. C., MacGillivray, H. T., Read,
  M. A., Tritton, S. B., Thomson, E. B., Kelly, B. D., Morgan, D. H., Smith,
  R. E., Driver, S. P., Williamson, J., Parker, Q. A., Hawkins, M. R. S.,
  Williams, P. M., Lawrence, A., 2001, MNRAS, 326, 1279
\bibitem[P\'econtal-Rousset et al. 2003]{fe03} P\'econtal-Rousset, A., et al., 2003, Euro3D Science Workshop, 21-23 May 2003, IoA, Cambridge, AN, , these proceedings.
\bibitem[Kelz et al. 2003]{ke03} Kelz, A., Roth, M.M., Becker, T., 2003, in Proc. SPIE, Vol. 4841, 1057
\bibitem[Kissler-Patig et al. 2003a]{kp03} Kissler-Patig, M., Copin, Y., Ferruit, P., P\'econtal-Rousset, A., Roth, M.M., 2003, Euro3D Data Format Definition, Euro3D Documentation.
\bibitem[Kissler-Patig et al. 2003b]{kp03b} Kissler-Patig, M., et al. 2003, Euro3D Science Workshop, 21-23 May 2003, IoA, Cambridge, AN, in preparation.
\bibitem[Le F\`evre et al. 2003]{lef03} Le F\`evre, 0. et al., 2003, in Proc. SPIE, Vol. 4841, 1670
\bibitem[Pecontal-Rousset et al. 2003]{pr03} P\'econtal-Rousset, A.,  Ferruit, P.,  Copin, Y., 2003, Euro3D I/O Libraries v1.0a, Installation Guide, Euro3D Documentation.
\bibitem[Roth et al. 2000]{ro00} Roth, M.M., Bauer, S., Dionies, F., et al., 2000, in Proc. SPIE, Vol. 4008, 277-288
\bibitem[S\'anchez et al. 2003]{san03} S\'anchez, S.F., Christensen, L.,
  Becker, T., Kelz, A., Jahnke, K., Benn, C.R., Garc\'\i a-Lorenzo, B., Roth,
  M.M., 2003, Euro3D Science Workshop, 21-23 May 2003, IoA, Cambridge, AN,
  these proceedings.
\bibitem[S\'anchez 2003]{san03b} S\'anchez, S.F.: 2003, Euro3D Science
  Workshop, 21-23 May 2003, IoA, Cambridge, AN, these proceedings.
\bibitem[Walsh \& Roth(2002)]{net02} Walsh, J.~R.~\& Roth, M.~M., 2002, The Messenger, 109, 54

\end{thebibliography}
\end{document}